\documentclass[fleqn,usenatbib]{mnras}

\usepackage{newtxtext,newtxmath}

\usepackage[T1]{fontenc}

\DeclareRobustCommand{\VAN}[3]{#2}
\let\VANthebibliography\thebibliography
\def\thebibliography{\DeclareRobustCommand{\VAN}[3]{##3}\VANthebibliography}

\usepackage{graphicx}	
\usepackage{amsmath}	
\usepackage{subcaption}
\usepackage{makecell}
\usepackage{hyperref}
\usepackage{footmisc}
\usepackage{orcidlink}
\usepackage[export]{adjustbox}
\usepackage{upgreek}

\usepackage{subcaption}
\usepackage{float}
\usepackage{graphicx}
\usepackage{caption}
\usepackage{lipsum}
\usepackage{makecell}
\usepackage{orcidlink}
\usepackage[export]{adjustbox}
\usepackage{ragged2e}
\usepackage{blindtext}

\usepackage{academicons}
\definecolor{orcidlogocol}{HTML}{A6CE39}
\newcommand{\sigmaRM}{$\sigma_{\rm RM}\,$}
\newcommand{\sigmapRM}{$\sigma^{\prime}_{\rm RM}$}
\newcommand{\RMunits}{rad$\,$m$^{-2}$}
\newcommand{\DMunits}{pc\,cm$^{-3}$}
\newcommand{\taus}{$\tau_s$}

\hypersetup{
    colorlinks=true,
    linkcolor=blue,
    filecolor=magenta,      
    urlcolor=magenta,
    pdfpagemode=FullScreen,
    }

\def\code#1{\texttt{#1}}

\title[ASKAP ICS FRBs depolarisation census]{A depolarisation census of ASKAP fast radio bursts}

\author[P. Uttarkar et al.]{\parbox{\textwidth}
{
Pavan A. Uttarkar$^{1}$\orcidlink{0000-0002-2346-6853}\thanks{E-mail: puttarkar@swin.edu.au},
Ryan~M.~Shannon$^{1}$\orcidlink{0000-0002-7285-6348},
Kelly~Gourdji$^{1,2}$\orcidlink{0000-0002-0152-1129},
Adam~T.~Deller$^{1}$\orcidlink{0000-0001-9434-3837},
Tyson Dial$^{1}$\orcidlink{0009-0004-1205-8805},\\
Marcin Glowacki$^{3}$\orcidlink{0000-0002-5067-8894},
Apurba Bera$^{4}$\orcidlink{0000-0002-2864-4110},
Alexa~C.~Gordon$^{5}$\orcidlink{0000-0002-5025-4645},
Stuart~D.~Ryder$^{6,7}$\orcidlink{0000-0003-4501-8100},
Nicolas Tejos$^{8}$\orcidlink{0000-0002-1883-4252},
Shivani~Bhandari$^{9}$\orcidlink{0000-0003-3460-506X},
Yuanming~Wang$^{1}$\orcidlink{0000-0002-0786-7307}
}
\\ \\
$^{1}$Centre for Astrophysics and Supercomputing, Swinburne University of Technology, Hawthorn, VIC\\
$^{2}$Australia Telescope National Facility, CSIRO, Space and Astronomy, PO Box 76, Epping, NSW 1710, Australia\\
$^{3}$Institute for Astronomy, University of Edinburgh, Royal Observatory, Edinburgh, EH9 3HJ, United Kingdom\\
$^{4}$International Centre for Radio Astronomy Research, Curtin University, Bentley, WA 6102, Australia\\
$^{5}$Center for Interdisciplinary Exploration and Research in Astrophysics (CIERA) and Department of Physics and Astronomy,\\ Northwestern University, Evanston, IL 60208, USA\\ 
$^{6}$School of Mathematical and Physical Sciences, Macquarie University, NSW 2109, Australia \\
$^{7}$Astrophysics and Space Technologies Research Centre, Macquarie University, Sydney, NSW 2109, Australia\\
$^{8}$Instituto de F\'isica, Pontificia Universidad Cat\'olica de Valpara\'iso, Casilla 4059, Valpara\'iso, Chile\\
$^{9}$SKA Observatory (SKAO), Science Operations Centre, CSIRO ARRC, Kensington WA 6151, Australia\\
}

\date{Accepted XXX. Received YYY; in original form ZZZ}

\pubyear{2015}

\begin{document}
\label{firstpage}
\pagerange{\pageref{firstpage}--\pageref{lastpage}}
\maketitle

\begin{abstract}
Fast radio bursts (FRBs) are luminous, dispersed pulses of extra-galactic origin. The physics of the emission mechanism, the progenitor environment, and their origin are unclear. 
Some repeating FRBs are observed to have frequency-dependent exponential suppression in linear polarisation fraction. This has been attributed to multipath propagation in a surrounding complex magneto-ionic environment.
The magnitude of depolarisation can be quantified using the parameter \sigmapRM, which can be used to model the magneto-ionic complexity of the medium. In addition to depolarisation, some repeating sources (in particular those with active magneto-ionic environments) have been identified to have co-located persistent radio sources (PRS).
Searches for depolarisation of non-repeating sources are challenging  due to the limited bandwidth of most FRB detection systems used to detect one-off bursts. However, even with a limited bandwidth, such depolarisation can be identified if it lies within the \sigmapRM\ sensitivity window of the telescope. 
In this paper, we present a search for depolarisation in $12$ one-off FRBs detected by the Australian SKA Pathfinder.
We report on the first strongly depolarised FRB detected by ASKAP (FRB~20230526A) and a marginal detection of depolarisation in a second.
We also report constraints on the presence of  a PRS coincident with FRB~20230526A using observations obtained with the Australia Telescope Compact Array. 
We use this to study the relationship between \sigmapRM\ and PRS luminosity. Our investigation supports a scenario in which repeaters and non-repeaters share a common origin and where non-repeaters represent an older population relative to repeating FRBs.


\end{abstract}

\begin{keywords}
fast radio bursts -- polarisation -- techniques: polarimetric
\end{keywords}



%
%

\section{Introduction}

Fast radio bursts (FRBs) are energetic transient  pulses that were first discovered by \cite{Lorimer:2007}. The first FRB was discovered in an archival dataset from observations with the 20-cm multibeam receiver on Murriyang, CSIRO’s Parkes radio telescope. The discovery of the first FRB and subsequent confirmation of this new source class by \cite{Thornton:2013} led to the opening up of an entirely new domain in transient astronomy. Since their discovery $\sim$2 decades ago, $\sim$800 FRBs have been reported \cite[e.g.,][]{Chime/FRBCollaboration:2020, Shannon:2024}, and a broad distinction in the population has emerged: those that repeat and those that apparently do not.  However, the mystery of the origin of these elusive bursts, the physical mechanism that produces the emission, and progenitor object class(es) has only deepened further, despite a relatively large sample set of bursts.

The relationship between repeating and non-repeating FRBs is still unclear. Repeating and non-repeating FRBs have been seen to have distinct spectro-temporal and polarisation properties. Many repeating FRBs show a distinct spectral drift across time, referred to as "sad-trombone" structure \citep{Hessels:2019}:  emission first appears at high frequency and then drifts to lower frequency over many milliseconds. They also show band-limited emission \citep{Hessels:2019}. This is in contrast to non-repeaters which have relatively less complex spectral behaviour \citep[e.g.,][]{Pleunis:2021}. Such spectral properties are extremely useful for inferring fundamental physical properties of FRB progenitors, in particular for understanding emission mechanism physics and the  surrounding physical properties of progenitors \cite[][]{Cordes:2017}.

In addition to spectral properties, repeaters and non-repeaters show a marked difference in polarisation. Some extremely active repeating FRB sources have been seen to have a large and highly variable rotation measure \citep[RM; e.g.,][]{Michilli:2018, hilmarsson21, Anna-Thomas:2023}, indicative of an extreme and complex magneto-ionic environment. The RM encapsulates the mean parallel component of magnetic field strength ($B_{||}$), weighted by electron column density along the line-of-sight of the observation. Repeaters such as FRBs~20121102A and 20190520B have been seen to have RMs similar to the values seen towards the centre of the Milky Way \cite[][]{Eatough:2013,Shannon:2013, Michilli:2018, hilmarsson21, Anna-Thomas:2023}. The polarimetric properties of FRBs have a unique imprint from this local complex environment. Hence, the polarisation properties of FRBs can be an ideal tool to study the progenitor circumburst environment. 

Some repeating FRB sources also show spectral depolarisation - a frequency-dependent decrease in the linear polarisation fraction \citep{Feng:2022}. Similar depolarisation behaviour has previously been observed in extended radio sources, such as the Crab nebula \citep{Burn:1966}. The depolarisation in the Crab nebula is attributed to the filamentary nature of its shell \citep{Burn:1966}. The magnitude of depolarisation is linked to the turbulence in the surrounding medium. 
A more turbulent medium will lead to a higher magnitude of depolarisation. This is because the medium will scatter the radiation through larger angles, resulting in it travelling though a larger volume through which the magnetic field is more perturbed \citep{Feng:2022}.
The magnitude of depolarisation is quantified  by \sigmaRM \citep{Burn:1966} or \sigmapRM  \citep[if we assume the emission is not intrinsically 100\% polarised;][]{Uttarkar:2023}. The repeating sources FRBs 20121102A and 20190520B notably have some of the largest magnitudes of \sigmaRM. They are observed to be associated with compact radio continuum sources termed persistent radio sources (PRS) \citep{Chatterjee:2017, Niu:2022, Bruni:2024}.  Superluminous supernovae have been observed to have similar properties to PRS \citep[e.g.,][]{Eftekhari:2019}, providing a potential plasma to depolarise FRB emission.   Given the limited sample size, the connection between the presence of a PRS and depolarisation is unclear.

Repeating FRB sources are ideally suited for searches for depolarisation due to the ability to schedule follow-up observations  over a range of frequencies (and potentially with multiple telescopes). Similar studies with (apparent) non-repeaters are usually difficult due to the limited bandwidth of commensal systems available for FRB detection, and the non-repeating nature of the sources. For example, with a limited bandwidth at a given central frequency, low values of \sigmapRM\ make an immeasurable impact, as the linear polarisation remains effectively unattenuated across the band, while high values of \sigmapRM\ are similarly indiscernible as the linear polarisation fraction is reduced to effectively zero across the band. 
However, even with a small available bandwidth, an instrument can confidently measure \sigmaRM for values for which the instrument is most sensitive; in general, when the depolarisation is $\sim$50\% near the centre of the band. The Australian SKA Pathfinder \citep[ASKAP;][]{Hotan:2021}, when observing over a bandwidth of $336$\,MHz, is most sensitive to \sigmapRM\ values of $\sim$12 \RMunits\ and $\sim$8 \RMunits\ at central frequencies of 1271~MHz and 824~MHz, respectively, \citep{Uttarkar:2023}.

There have only been limited searches for depolarisation in apparently non-repeating FRB sources. The spectro-polarimetric analysis of 25 apparent non-repeaters by \cite{Shreman:2024} report \sigmaRM\ for FRBs 20220914A, 20220926A, and 20221027A to be between 0.3--3 \RMunits, and disfavour the presence of inhomogeneous highly magnetised plasma near the vicinity of their progenitors, based on the observed RM and the measured \sigmaRM\ of their sources. 
Depolarisation studies with the Canadian Hydrogen Intensity Mapping Experiment (CHIME) reported only one FRB with a marginal depolarisation of 20\% across the band \citep[FRB~20190217A;][]{Pandhi:2024}. 
This FRB was detected when CHIME was unable to confidently associate most FRBs with a host galaxy.
In contrast, ASKAP has been regularly associating FRBs with host galaxies for more than half a decade \cite[][]{Bannister:2019,Shannon:2024}. 
An initial study of a subsample of FRBs from ASKAP \cite[][]{Day:2020} did not show any significant depolarisation \citep{Uttarkar:2023}. However, it is unclear if the non-detection of exponential suppression of depolarisation is purely instrumental selection effect or a bona-fide characteristic of non-repeaters \citep{Uttarkar:2023}.

The incoherent sum detection system (ICS) on ASKAP searches the incoherent addition of intensities from ASKAP antennas for FRB detection. The ICS sytem uses a GPU-based search pipeline to detect FRBs \citep{Shannon:2024}. Any detection of an FRB candidate triggers a voltage download from individual antennas. These voltages are further used to  localise, beamform and derive polarimetric and high time resolution properties of FRBs \citep{Scott:2023}. Due to the incoherent nature of the detection system and the coherent nature of the localised and beamformed data, the latter will have a greater signal-to-noise ratio (S/N). 

In this paper, we extend the depolarisation analysis reported in \cite{Uttarkar:2023} to an additional $12$ FRBs. This includes the discovery of strong depolarisation in the (apparently) non-repeating FRB~20230526A. In Section \ref{sec:methods}, we report the methods we use for the depolarisation analysis. 
We report our results from the depolarisation analysis in Section \ref{sec:results}. In this section we also report on searches for repetitions and a counterpart persistent radio source for FRB~20230526A. We discuss the implication of our results in Section \ref{sec:discussion}. We conclude in Section \ref{sec:conslusion}.

\section{Observation and methodology}

\label{sec:methods}

\subsection{ASKAP-ICS FRB sample set}

We use a subset of FRBs reported by \cite{Shannon:2024} for our analysis,   focusing on all FRBs that had high quality polarimetric data detected before 19 July 2023.
We refer to \cite{Bannister:2019} and \cite{Shannon:2024} for a detailed description of the detection pipeline and localisation system. The list of FRBs in our sample and their properties are listed in Table \ref{tab:comparision_table}. 
We use  high-time resolution (HTR) tied-array beam data sets from Scott et al. 2025 (in prep) for our analysis. Dynamic spectra in Stokes I, Q, U, and V were created using the HTR complex voltages. The frequency resolution used to perform depolarisation analysis and the scattering timescales used in the study are also listed in Table \ref{tab:comparision_table}. 

\subsection{Assessing depolarisation properties}
A linearly polarised radio pulse travelling through a cold magnetised medium experiences a phase shift in the polarisation position angle (PPA) due to the magnetic field component parallel to the line-of-sight to the observer. 
This is due to the circular birefringence of the medium: left and right hand circularly polarised waves travel at different speeds through magnetised plasma.
The rotation of the PPA is dependent on the wavelength of the radiation, and it is defined as 
\begin{equation}
    \psi = \psi_{\rm o} + \rm RM\,(\lambda^2 - \lambda_{o}^2),
    \label{eq:PPA_quadratic}
\end{equation}
where RM is the observed rotation measure, $\lambda_{\rm o}$ is the reference wavelength, $\lambda$ is the wavelength, and $\psi_{\rm o}$ is the intrinsic PPA. The RM contributed by the cold magnetised plasma can be defined using physical quantities as
\begin{equation}
    {\rm RM} = \frac{e^3}{2\pi m_e^2 c^4}\int_{0}^{d}\frac{n_e B_{||}}{(1+z)^2} dl,
	\label{eq:RM_calulation}
\end{equation}
where $e$ is the charge of the electron, $m_e$ is the mass of the electron, $n_e$ is the density of charged particles, $c$ is the speed of light, $z$ is the redshift of the plasma, $B_{||}$ is the magnetic field component parallel to the line of the sight to the observer, and $d$ is the distance to the source. 

To assess the depolarisation properties, we measure the linear polarisation fraction of the burst after de-rotating the Stokes spectra to account for Faraday rotation using
\begin{equation}
\begin{bmatrix}
Q_{\rm de-rot} \\
U_{\rm de-rot} \\
\end{bmatrix}
 = 
\begin{bmatrix}
\cos 2\psi & \sin 2\psi \\
-\sin 2\psi & \cos 2\psi \\
\end{bmatrix}
\begin{bmatrix}
Q \\
U \\
\end{bmatrix},
\end{equation}
where $Q_{\rm de-rot}$ and $U_{\rm de-rot}$ are the Faraday de-rotated spectra and $\psi$ is the PPA. We use the RMs reported by Scott et al. 2025 (in prep), to de-rotate the Stokes-Q and U spectra. These are listed in Table \ref{tab:comparision_table}.

We maximise the polarised SNR in our spectra using the match filtering methods described in Sections 2.3 and 2.4 of \cite{Uttarkar:2023}. We use a single-component exponentially convolved Gaussian kernel on each of the bursts to create a matched filter to maximise the SNR when averaging the Stokes Parameters over the off pulse window. The linear polarisation fraction for each of the bursts is $L=\sqrt{Q_{\rm de-rot}^2+U_{\rm de-rot}^2}$. Following \cite{Everett:2000}, we further de-bias the linear polarisation fraction using
\begin{equation}
    L_{\rm de-bias}  =  
    \begin{cases}
        \sigma_{I}\,\sqrt{\left(\frac{L}{\sigma_{I}}\right)^2 - 1}\,,& \text{if } \frac{L}{\sigma_I}\geq 1.57\beta \\
    0,              & \text{otherwise},
    \end{cases}
	\label{eq:debais}
\end{equation}
where $\sigma_L$ is the uncertainty in linear polarisation fraction, $L$ is the linear polarisation fraction, $\sigma_I$ is the Stokes-I uncertainty, and $\beta$ is the polarised SNR threshold. We do not include the linear polarisation measurements of frequency channels that do not satisfy the above criteria when assessing the depolarisation model.

The uncertainties on the linear polarisation fraction are estimated using
\begin{equation}
    \sigma_{L/I}(\nu) =  \frac{\sqrt{\left(\sigma_L^2\,P_I^2 + L_{\rm de-bias}^2\,\sigma_I^2 \right)}}{P_I^2}, 
    \label{eq:L_I_error}
\end{equation}
where $\sigma_L$ is the linear polarisation fraction uncertainty calculated using Equation 9 in \cite{Uttarkar:2023}, $P_I$ is the SNR maximised on-pulse time averaged Stokes-I component, and $L_{\rm de-bias}$ is the de-biased linear polarisation fraction.

\subsubsection{Modelling FRB depolarisation}
We consider two independent models for the linear polarisation of the bursts. In the first (model 1), we assume that there is presence of depolarisation due to multipath propagation of radio waves, described by \cite{Burn:1966}. In the second (model 2), we assume a constant fractional polarisation model across the band.
We refer to \cite{Uttarkar:2023} for a detailed description of the models.

We further bifurcate the first model into two different sub-models to better capture the possible depolarisation behaviour of the bursts. The first (model 1(a)) assumes only Burn's law of depolarisation due to multipath propagation. The second (model 1(b)) is a modified version of Burn's law that accounts for potential  lower intrinsic polarisation fraction from the source. The three models are thus
\begin{equation}
    \begin{split}
        &P_{\rm Burn}(\lambda) =  \exp{\left(-2\sigma_{\rm RM}^2\lambda^4\right)},\\
        &P_{\rm mod-Burn}(\lambda) =  P_i\exp{\left(-2\sigma^{\prime2}_{\rm RM}\lambda^4\right)}, {\rm and}\\
        &P_{\rm const}(\lambda)   = P_i,   
    \end{split}
    \label{eq:Burns_law}
\end{equation}
where \sigmaRM and \sigmapRM\ encapsulate the magnitude of depolarisation for the FRB, $\lambda$ is the wavelength, and $P_i$ is the polarisation fraction intrinsic to the source. We use a Bayesian approach to estimate the model preference as described in \cite{Uttarkar:2023}. We use the log$_{10}$ Bayes evidence to identify whether there is a preferred model. Following \cite{Trotta:2008}, we consider a log$_{10}$ evidence greater than $10$ to be a strong preference for a model.

\begin{figure*}
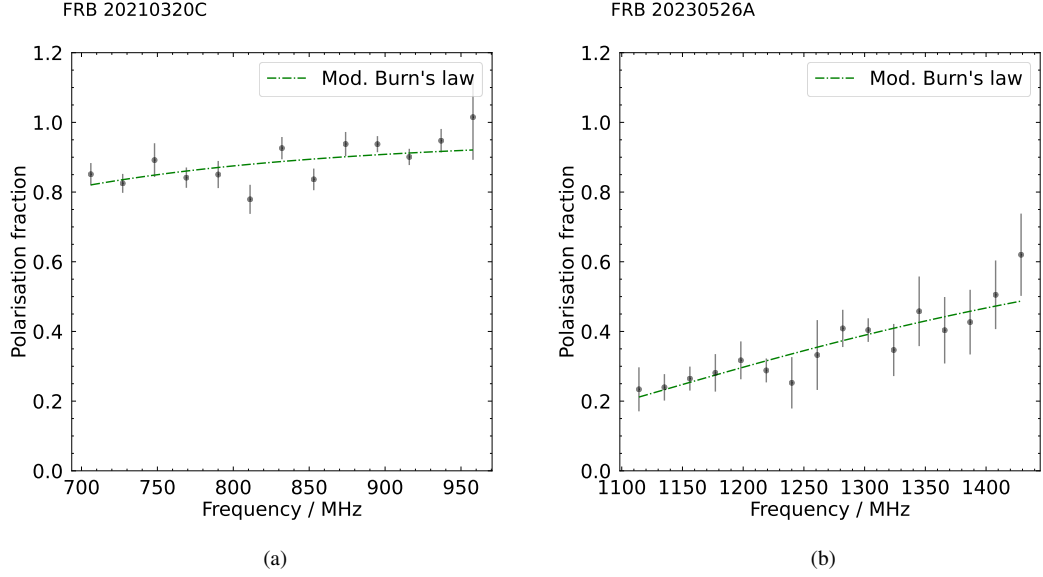

    \begin{subfigure}[t]{.4\linewidth}
        \centering\includegraphics[width=1\linewidth]{DEPOLARISATION_PROFILES_NEWDATA/20210320A.pdf}
        \caption{}
    \end{subfigure}  
    \begin{subfigure}[t]{.4\linewidth}
        \centering\includegraphics[width=1\linewidth]{DEPOLARISATION_PROFILES_NEWDATA/20230526A.pdf}
        \caption{}
    \end{subfigure}  
    \caption{Spectral depolarisation of FRBs 20210320C and 20230526A. 
    The most favoured model is plotted for each burst by the dashed line. FRBs 20210320C and 20230526A show evidence for exponential suppression of the linear polarisation fraction over the ASKAP observing band.}
    \label{fig:depolarisation_model}
\end{figure*}

\begin{figure*}
    \begin{subfigure}[t]{.45\linewidth}
        \centering\includegraphics[width=1\linewidth]{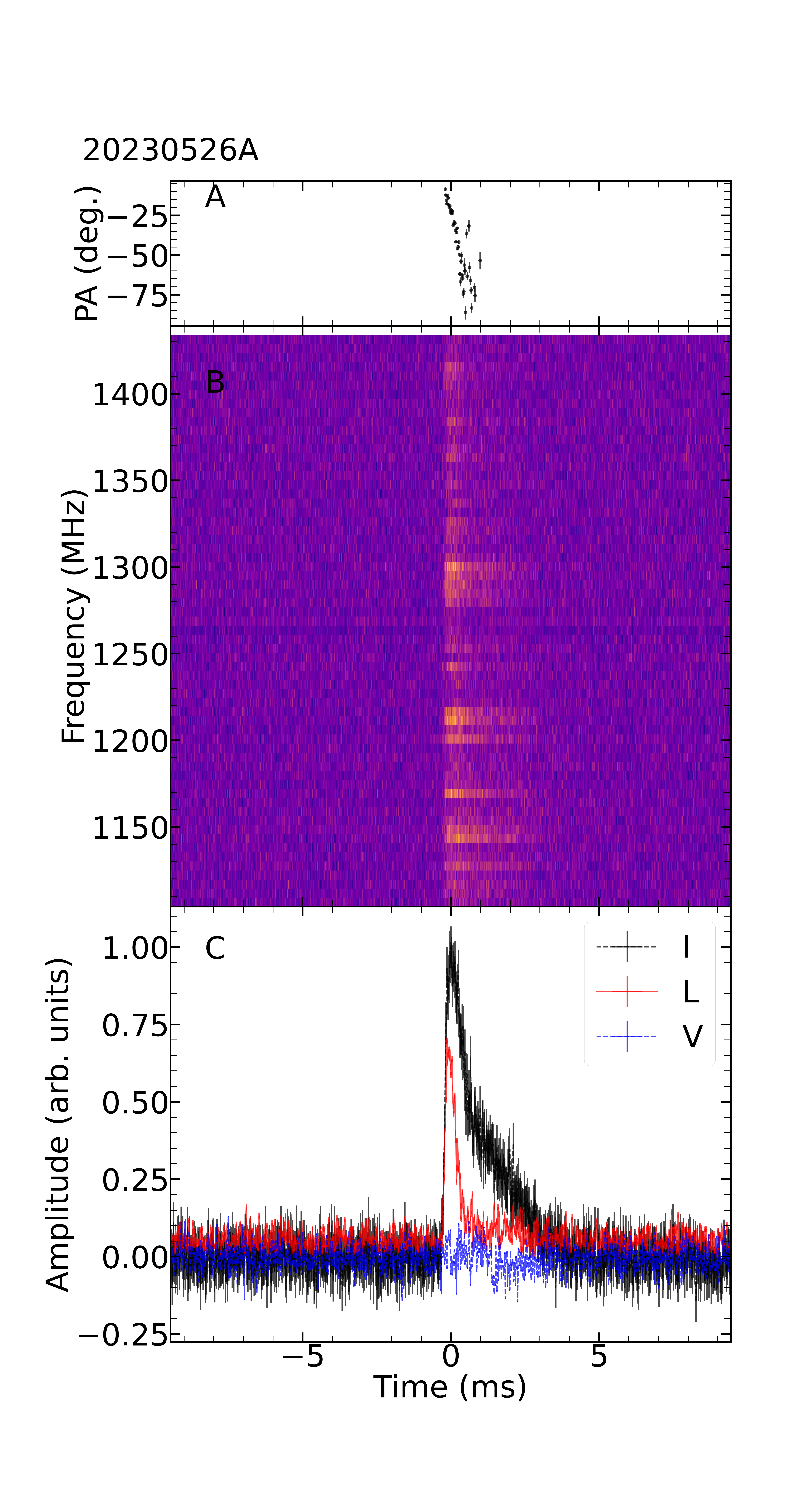}\caption{}
    \end{subfigure}
    \begin{subfigure}[t]{.45\linewidth}
        \centering\includegraphics[width=1\linewidth]{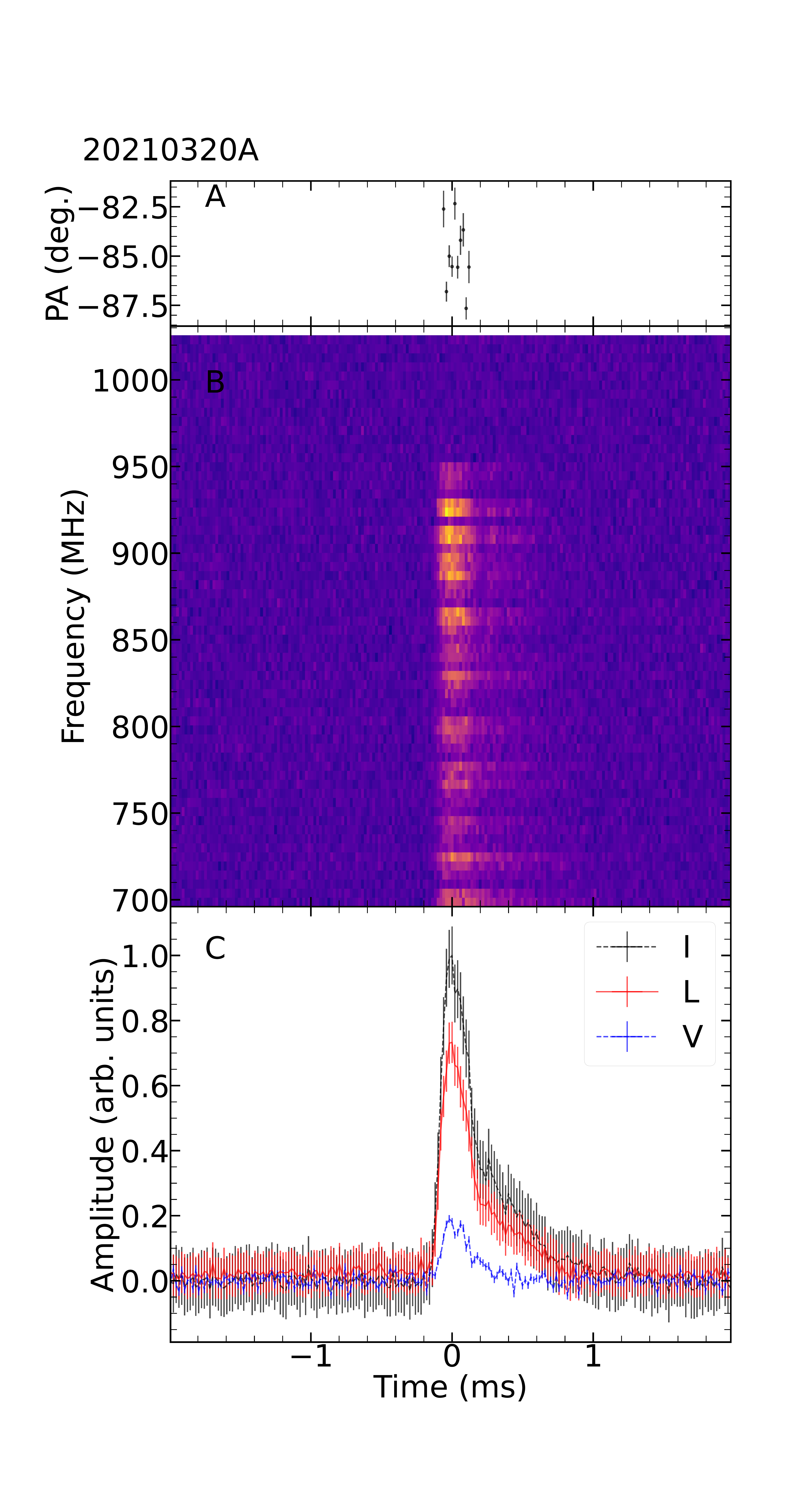}\caption{}
        \end{subfigure}
    \caption{Dynamic spectra of FRBs 20210320C and 20230526A. Panels A, B, and C show the PPA, dynamic Stokes-I spectra, and the frequency averaged polarisation profiles for both FRBs, respectively. The frequency averaged total intensity, linear, and circular polarisation fraction are shown in black, red, and blue lines, respectively. We show the bursts with a spectral and temporal resolution of $5.25$ MHz and $20$ $\upmu$s, respectively.}
    \label{fig:Stokes_I_spectra}
\end{figure*}

\begin{table*}
	\begin{center}\renewcommand{\arraystretch}{1.5}%
		\begin{tabular}{ c c c c c c c c c c c c c c}
			\hline
			\thead{FRB \\ (TNS)}  & \thead{DM$^\dag$ \\ (pc cm$^{-3}$)} & \thead{RM$^\ddag$ \\ (rad m$^{-2}$)} & z$^\dag$ & \thead{\sigmapRM$^{\nmid}$ \\ (rad m$^{-2}$)} & P$_{i}$$^{\nmid}$ & \thead{\sigmaRM$^{\nmid}$ \\ (rad m$^{-2}$)} & \thead{$\nu_{\rm centre}$$^{\dag}$ \\ (MHz)} & \thead{$\delta\nu$$^{\nmid}$ \\ (MHz)} & \thead{$\tau$$^{\nmid}$ \\ (ms)} & \thead{Bayes Factor$^{\nmid}$ \\ (log$_{10}$)}\\
			\hline

			20191228A & 297.5(5) & 11.3$\pm$0.8 & 0.243 & $<$5.17 & 0.93$^{+0.05}_{-0.06}$ & $<$3.93 & 1272 & 20.978 & 5.69$^{+0.15}_{-0.14}$ & -0.6 \\
			20200430A$^{\rm a}$ & 380.1(2) & 195.3$\pm$0.6 & 0.161 & $<$4.41 & 0.52$^{+0.19}_{-0.09}$ & $<$3.51 & 864 & 20.978 & 7.27$^{+0.2}_{-0.21}$ & -0.7 \\
			20210117A & 730(1) & -45.4$\pm$0.7 & 0.214 & $<$3.05 & 0.97$^{+0.02}_{-0.02}$ & $<$2.25 & 1272 & 20.978 & 0.09$^{+0.07}_{-0.06}$ & -0.9 \\
			20210320C & 398.3(7) & 288.8$\pm$0.2 & 0.280 & 1.83$^{+0.07}_{-0.08}$ & 0.97$^{+0.02}_{-0.02}$ & 1.59$^{+0.17}_{-0.24}$ & 864 & 20.978 & 0.38$^{+0.2}_{-0.18}$ & 1.5 \\
			20210407E & 1785.3(3) & -8.9$\pm$0.5 & no host & $<$2.53 & 0.96$^{+0.02}_{-0.01}$ & $<$1.98 & 1272 & 20.978 & 0.27$^{+0.04}_{-0.15}$ & -1 \\
			20210912A & 1234.5(2) & 5.7$\pm$0.4 & no host & $<$5.06 & 0.61$^{+0.02}_{-0.02}$ & $<$4.3 & 1272 & 20.978 & 0.45$^{+0.02}_{-0.02}$ & -0.1 \\
			20220501C & 449.5(2) & 35.2$\pm$0.4 & 0.381 & $<$2.07 & 0.64$^{+0.04}_{-0.04}$ & $<$1.38 & 864 & 20.978 & - & -1.2 \\
			20220610A & 1458.1(2) & 217.0$\pm$2 & 1.016 & $<$1.7 & 0.99$^{+0.01}_{-0.01}$ & $<$1.19 & 1272 & 20.978 & 0.54$^{+0.03}_{-0.05}$ & -1.1 \\
			20220725A & 290.4(3) & -26.0$\pm$2.0 & 0.193 & $<$4.13 & 0.75$^{+0.13}_{-0.09}$ & $<$3.22 & 920 & 20.978 & 2.4$^{+0.12}_{-0.12}$ & -0.7 \\
			20221106A & 343.8(8) & 445.0$\pm$1 & 0.204 & $<$10.04 & 0.86$^{+0.07}_{-0.04}$ & $<$8.38 & 1632 & 20.978 & - & -0.3 \\
			20230526A & 361.4(2) & 613.0$\pm$2.0 & 0.157 & 12.62$^{+0.25}_{-0.23}$ & 0.79$^{+0.12}_{-0.11}$ & 11.23$^{+0.83}_{-1.04}$ & 1272 & 20.978 & 1.39$^{+0.04}_{-0.04}$ & 5.7 \\
			20230718A & 477.0(5) & 243$\pm$1.0 & 0.035 & $<$5.91 & 0.91$^{+0.05}_{-0.06}$ & $<$5.38 & 1272 & 20.978 & - & 0.7 \\
			\hline
		\end{tabular}
	\end{center}
        \caption{Depolarisation properties of the ICS FRB sample. The FRB TNS name, DM, RM, host redshift, \sigmapRM, P$_{i}$, \sigmaRM, centre frequency , the frequency resolution used for depolarisation model fit $\delta \nu$, and the scattering timescale $\tau$ for the ICS FRB sample are reported here. We report the log$_{10}$ Bayes Factor (BF) for the modified Burn's law and the constant depolarisation models. The higher positive value indicates higher preference to the modified Burn's law. The values for columns marked with $^\dag$ are parameters from \protect\cite{Shannon:2024}. The columns marked with $^{\nmid}$ are the parameters derived in this work. $\ddag$RMs for the bursts are derived from Scott et al. 2025 (in prep). We use simple boxcar for FRBs where no scattering timescales are provided. $^{\rm a}$  We use the data for FRB 20200430A without any polarisation calibration due to insufficient S/N of the calibration pulsar J2045$-$1616.}
	\label{tab:comparision_table}
\end{table*}

\begin{figure*}
        \centering\includegraphics[width=1\linewidth]{RM_TAU_SIGMA/RM_tau_sigma_RM.pdf}
        \caption{The RM, \sigmapRM, and \taus\ relationships.   The top panel shows the relationship between \taus\ and \sigmapRM. The relationship between RM and \sigmapRM\ is shown in the bottom panel. The repeating FRB sources from \protect\cite{Feng:2022} are displayed as  grey crosses. The sample of FRBs from \protect\cite{Uttarkar:2023} are shown as grey triangles. This latter sample includes the repeating FRB source FRB~20190711A. The sample of ICS FRBs from this work are shown in blue circles and green stars. The blue line is the (logarithmic) linear correlation to the trend for the repeating FRB dataset from \protect\cite{Feng:2022}, and the red shaded region shows the 1-$\sigma$ error region of the correlation. The upper limits on the measurements of the \sigmapRM\ are shown for the data points with arrows. The error bars on the RM measurements are smaller than the markers.}
    \label{fig:correlation}
\end{figure*}



\begin{table*}
    \begin{center}\renewcommand{\arraystretch}{1.5}
    \begin{tabular}{c c c c}
        \hline
        FRB TNS &  \thead{PRS luminosity\\(10$^{29}$ ergs s$^{-1}$ Hz$^{-1}$)} & \sigmaRM\ & \thead{RM\\($\rm rad\,m^{-2}$)} \\
        \hline
        FRB~20121102A & 2.1$\pm$0.044$^{\ddag}$ & 30.9$\pm$0.4$^{*}$ & $\sim$10$^5$\\
        FRB~20180301A & $<$1.8$^{**}$          & 0.12$\pm$0.01$^{*}$ & -130$\pm$20\\
        FRB~20180916B & $<$0.0048$^{\uparrow}$          & 0.12$\pm$0.01$^{*}$ & -115$\pm$20\\        
        FRB~20190520B & 3.0$\pm$0.074$^{\nmid}$  & 218.9$\pm$10.2$^{*}$ & $\sim-$2.4$\times$10$^{4}$ $-$ $\sim$1.2 $\times$10$^{4}$\\
        FRB~20191001A & <0.2$^{\downarrow}$           & 4.1$\pm$0.1$^{\downarrow}$ & 55.5$\pm$0.9\\ 
        FRB~20201124A & 0.053$\pm$0.006$^{\dag}$ & 2.5$\pm$0.1$^{*}$& 616.4$\pm$7.2\\
        FRB~20230526A & <0.11$^{\rm (this\,work)}$           & 12.61$\pm$0.2$^{\rm (this\,work)}$ & 613.0$\pm$2.0\\ 
        \hline
    \end{tabular}
    \caption{PRS luminosity and depolarisation properties of FRBs. We use sample of FRB for which \protect\sigmaRM measurements and PRS follow-up information is available. The PRS luminosities and \protect\sigmaRM are derived from $^{\ddag}$\protect\cite{Chatterjee:2017}, $^{\downarrow}$\protect\cite{Bhandari:2020}, $^{\uparrow}$\protect\cite{Marcote:2020}, $^{**}$\protect\cite{Bhandari:2022}, 
    $^{*}$\protect\cite{Feng:2022}, 
    $^{\nmid}$\protect\cite{Niu:2022}, and $^{\dag}$\protect\cite{Bruni:2024}. The RM values for FRBs 20121102A, 20180916B, 20191001A, 20190520B, and 20201124A are derived from \protect\cite{Michilli:2018}, \protect\cite{Feng:2022}, \protect\cite{Bhandari:2020},  \protect\cite{Anna-Thomas:2023}, and \protect\cite{Kumar:2022}, respectively. The RM value of FRBs 20180301A and 20201124A are the average of all RM values reported by \protect\cite{Kumar:2022} and \protect\cite{Kumar:2023}, respectively.}
    \label{tab:PRS_sigma_RM}
    \end{center}
\end{table*}

\section{Results}
\label{sec:results}

In our sample of $12$ FRBs, we place upper limits for $10$ FRBs for depolarisation, while for the other two, we have found modest to significant evidence for depolarisation. 
We show the linear polarisation fraction across the frequency band for the two FRBs that show evidence for depolarisation, FRBs 20210320C and 20230526A, in Figure \ref{fig:depolarisation_model}. Spectropolarimetry for all FRBs are shown in Figure \ref{fig:depolarisation_model_apdx} in the appendix. The model fits for the preferred depolarisation model are shown in dashed lines. Burn's law, modified Burn's law, and the constant depolarisation models are shown in blue, green, and red dashed lines, respectively. We show the 95\% upper limit on the modified Burn's law in a solid black line for all the FRBs, except those for which we have measurements of \sigmapRM. 


FRB~20230526A shows a clear suppression of linear polarisation fraction across the ASKAP band (see Figure \ref{fig:depolarisation_model}(b)).  
It has a Bayes factor (BF) of $\log_{10}$=5.7  in support of the depolarisation model over those that do not. This corresponds to modest to substantial evidence towards the spectral depolarisation model \citep{Trotta:2008}.  We also note that FRB~20230526A has the largest magnitude of |RM|  (613 \RMunits) in our sample. 
We also find modest evidence for depolarisation in FRB~20210320C. The frequency-averaged polarisation spectrum, PA angles, and Stokes-I dynamic spectrum for FRBs~20210320C and 20230526A are shown in Figure \ref{fig:Stokes_I_spectra}.
We note that  FRB~20230526A  shows temporal variation in PA across the burst.  While this will cause modest depolarisation when averaging individual channels over the pulse profile, it will not induce spectral depolarisation.    

We place upper limits on the magnitude of depolarisation for the remaining  FRBs that range from $1.7$ \RMunits\ to $10.2$ \RMunits. 

\begin{figure}
    \centering
    \includegraphics[width=1\linewidth]{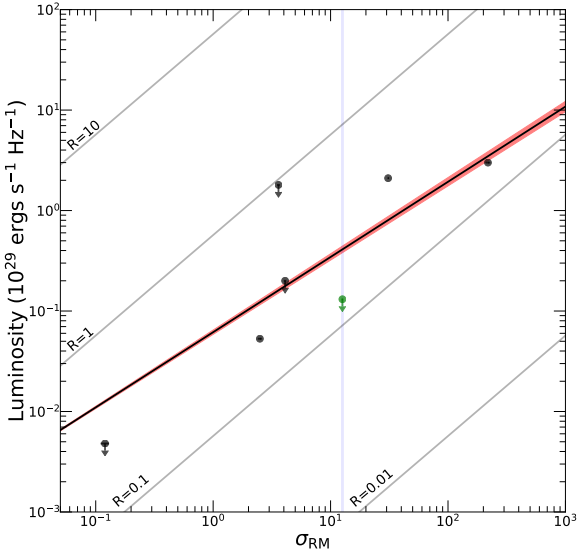}
    \caption{The PRS Luminosity--\protect\sigmaRM\ relationship. Repeating FRBs with detected PRS are shown in grey points. FRB~20230526A is shown as a green point. The upper limits on the PRS emission are shown with downward-facing arrows. The power law-fit to the data is shown as a black curve, and the 1-$\sigma$ error range in the red shaded region. The 1-$\sigma$ error bar on the \protect\sigmaRM\ measurement for FRB~20230526A is shown in the blue highlighted region. We show the radius scaling from \protect\cite{Yang:2022} from Equation \protect\ref{eq:luminosity_sigma_RM} in gray lines.  The PRS luminosities used are listed in  Table \protect\ref{tab:PRS_sigma_RM}. }
    \label{fig:PRS_sigma_RM}
\end{figure}

\subsection{Follow up of FRB~20230526A}

As we detected significant depolarisation from  FRB~20230526A, we conducted further observations to determine if the host resembles those of other repeating FRBs, if the FRB is a repeater, and if there is a PRS co-located with the FRB position. 

\subsubsection{Host galaxy properties}

The detection of the source from ASKAP and subsequent voltage capture enabled the localisation of the FRB to Right Ascension (RA) and Declination (Dec) of 01h28m55.77s  and $-$52d43m02.46s, respectively \cite[][]{Shannon:2024}.  The position uncertainties are $140$\,mas and $115$\,mas in RA and Dec, respectively.
The subarcsecond localisation enabled the confident association of FRB~20230526A with a host galaxy not atypical of other host galaxies of ASKAP-localised FRBs \cite[][]{Bhandari:2020,Heintz:2021,Gordon:2023}.
The position of FRB~20230526A is coincident with a galaxy identified in the Dark Energy Survey \cite[][]{DECALS}. 
Following the localisation of FRB~20230526A we conducted follow up imaging observations with the ESO Very Large Telescope (VLT) using FORS-2 (optical) and HAWK-I (near-infrared), as well as X-shooter for spectroscopy. This enabled \cite{Shannon:2024} to associate FRB~20230526A with its host galaxy at a redshift $z=0.1570$, with the association probability estimated to have 99.7\% confidence, based on the Probabilistic Association of Transients to their Hosts framework \cite[PATH;][]{aggarwa:2021}.  Gordon et al. (in prep) study the morphology and properties of the largest sample of ASKAP-localised FRBs.  They find that the host galaxy of FRB~20230526A has a stellar mass of $\sim10^{10.73}\,M_\odot$ and a light profile consistent with a late-type disk galaxy, comparable to the majority of the ASKAP sample. This is in contrast to the dwarf-galaxy hosts of FRB~20121102A \cite[][]{Tendulkar:2017} and FRB~20190520B \cite[][]{Niu:2022}.

\subsubsection{Murriyang UWL searches for repetitions}

We searched for repeat bursts from FRB~20230526A using the ultrawideband low \cite[UWL;][]{Hobbs:2019} receiver on the Murriyang radio telescope. The source was observed for a total of 6.43 hours spread across four epochs between 16 Dec 2023 to 24 Jan 2024. The data were recorded in the pulsar search mode with coherent de-dispersion at 315.82 \DMunits, spectral and temporal resolution of 0.5 MHz and 64\,$\upmu$s, respectively. 
We searched the data using a sub-banded search pipeline consisting of 
\code{heimdall} \citep{Barsdell:2012} and \code{fetch} \citep{Agarwal:2019}.  A detailed description of the search pipeline can be found in  \cite{Kumar:2021}. We searched the dataset for a DM range from 100 \DMunits\ to 2100 \DMunits, with 13 sub-bands. We searched for single pulses in sub-bands of 64~MHz, 128~MHz, 256~MHz, 512~MHz, 832~MHz, 1664~MHz, and~3328 MHz bandwidth. The single pulse searches included an overlapping sub-band search to avoid missing any candidates present at band edges. We did not find any promising candidates from the observations above a S/N of 7.5. Our fluence completeness threshold is 2 Jy ms for bursts of width 30\,ms. 

\subsubsection{ATCA searches for a PRS}

We searched for a PRS coincident with the position of the FRB~20230526A using the Australia Telescope Compact Array (ATCA) using the C/X receiver system, in two bands  centered at 5500 MHz and 7500 MHz, each with a bandwidth of 2048 MHz. We observed the source with a spectral resolution of 1 MHz, with a total integration time of 10.52 hours. The observations were conducted on the 15th of February 2024 (UTC) in \code{6A} configuration\footnote{\url{https://www.narrabri.atnf.csiro.au/operations/array_configurations/}}, which has a maximum baseline length of 6\,km. The observation was conducted using the Compact Array Broad-band Backend  \cite[CABB, ][]{Wilson:2011}.  We used the radio galaxies  PKS~1934$-$638 and PKS~0131$-$522 for flux and phase calibration, respectively.

The correlated data from CABB were reduced using the \code{miriad} software package \citep{Sault:1995}. We performed the RFI flagging using \code{miriad} utilities \code{blflag} and \code{pgflag}. 
Primary calibration was applied using the \code{mfcal} routine, while  gain and phase calibration were undertaken using \code{gpcal}.
The calibrated data were then converted into a dirty map using \code{invert}, which was deconvolved using \code{mfclean}. We performed the deconvolution on individual intermediate frequencies of 5.5 GHz and 7.5 GHz.
The data were self-calibrated using the miriad \code{selfcal} utility. 


We did not detect any continuum emission co-located with the FRB localisation. We force-fitted a point source at the position of the ASKAP localisation, placing a 3-$\sigma$ upper-limit of 19 $\upmu$Jy and 21 $\upmu$Jy for the 7500 and 5500 MHz bands, respectively, on any point source associated with the FRB 20230526A. At redshift of 0.157, with a luminosity distance of 755.2 Mpc, and assuming a flat spectral index, we place a 3-$\sigma$ luminosity upper limit to be 0.11$\times$10$^{29}$ erg s$^{-1}$ Hz$^{-1}$.



\section{Discussion}
\label{sec:discussion}

We present the relationship between \sigmapRM, \taus, and RM in Figure \ref{fig:correlation} for both our sample of FRBs and those previously published in \citet{Feng:2022} and \citet{Uttarkar:2023}.
Both  FRBs in the new sample that show depolarisation are consistent within the 1-$\sigma$ uncertainty region of the power law for \sigmaRM\ and \taus, and \sigmaRM\ and RM reported by \cite{Feng:2022}. 
This suggests that RM, scattering and \sigmapRM\ are consistent with originating from the same medium. 




The repeating FRBs and (apparent) non-repeating FRBs have a large distribution in the spectral and polarimetric properties, e.g., large variations in  RM \citep{Cho:2020, Day:2020, Sand:2024}; magnitude of \sigmaRM\ \citep{Feng:2022, Uttarkar:2023}; and the presence of circular polarisation, making it difficult to definitely determine the  potential progenitor models. However, the growing sample of FRBs has slowly started showing broad indications of the surrounding environments of repeating and non-repeating FRB sources that could help in our understanding of FRB progenitors and the connection between repeating and non-repeating sources. 

PRS are speculated to be powered by relativistic electrons, accelerated by a young compact source \citep[e.g.][]{kashiyama17,Metzger:2017, Margalit:2018}. The complex magneto-ionic environments that harbour PRS are also believed to be responsible for the depolarisation seen in some of the repeating FRB sources \citep{Feng:2022}. However, similar depolarisation in non-repeaters has not been confidently observed until now. The origin of depolarisation in an apparently non-repeating source and how it relates to the repeating population is unclear. In this section, we investigate some such outstanding questions using the polarisation properties of FRBs and PRS measurements and upper limits.

\subsection{Origin of depolarisation}


The depolarisation analysis of repeaters by \cite{Feng:2022} and the study by \cite{Wan-Jin:2023} on the temporal variation of \sigmaRM\ for FRB~20201124A supports the argument that depolarisation likely occurs in the local magneto-ionic environment of the progenitor sources. 
However, our previous investigation of depolarisation behaviour of a set of ASKAP non-repeaters shows an inconsistent relationship between \sigmapRM and RM, and between \sigmapRM and \taus, compared to repeating FRB sources \citep{Uttarkar:2023}. 
This suggests a relatively less magnetised medium for the non-repeating FRB sources and that the RM and scattering could be originating in a different medium  \cite[e.g., host galaxy interstellar medium;][]{Uttarkar:2023}.
In our extended dataset, aside from FRBs 20210320C and 20230526A, we observe a similar behaviour with \cite{Uttarkar:2023}, where the correlation between the \sigmapRM and \taus\ for non-repeating sources with upper-limits is inconsistent with the repeater trend reported by \cite{Feng:2022} (see Figure \ref{fig:correlation}). 
This provides further evidence that for most non-repeating FRBs, scattering and RM originate in environments different to those that can cause depolarisation. 

The depolarisation characteristics of FRBs can be used to probe the surrounding magneto-ionic environment of the progenitor, such as the turbulence and magnetisation in the local magneto-ionic environment. In a model for the medium presented in \cite{Feng:2022}, in which the depolarisation occurs in a screen between the emitting source and the observer, the degree of magneto-ionic activity or magnetic energy density is parameterized by 
$\delta(n_e B_{\rm ||})$, which captures  fluctuations in both plasma density and magnetic field strength. 
Following Equation S9 of \cite{Feng:2022}, turbulence in a screen with depolarisation with magnitude  \sigmapRM\ will result in
\begin{equation}
      \delta(n_e B_{\rm ||}) =   \left(\frac{2 \pi m_e^2 c^4}{e^3}\right)\sigma^{\prime}_{\rm RM}\,R^{-1/2}\,l_{\rm screen}^{-1/2},
\end{equation}
where $l_{\rm screen}$ is the length scale of the plasma screen and $R$ is the radius of the progenitor's turbulent environment.
One possibility is that the PRS is a supernova remnant. 
While it is unclear what the properties are for this source (such as size), we can estimate properties based on the well-studied Galactic supernova remnant SN~1006 and include scaling relations that can be used to consider alternate scenarios.  
Thus, if we assume the supernova remnant is similar to that of SN 1006 that has a shell radius of 21 pc \citep{Frail:1995}, and assuming l$_{\rm screen}$ and \sigmapRM to be 10$^{15}$ cm and 12 \RMunits\, respectively, we find 
\begin{equation}
      \delta(n_e B_{\rm ||}) =  0.2 \times10^3 \rm \upmu G\,cm^{-3}\left(\frac{\sigma^{\prime}_{\rm RM}}{12~\rm rad\,m^{-2}}\right)\,\left(\frac{R}{21~\rm pc}\right)^{-1/2}\,\left(\frac{l_{\rm screen}}{10^{15} cm}\right)^{-1/2}.
\end{equation} 

Similarly, if we consider younger supernova remnants which have smaller radio shell diameters by a factor of $\sim$100-1000 relative to SN1006, it will result in $\delta(n_e B_{\rm ||})$ values to be $\sim2\times10^3\rm \upmu G\,cm^{-3}$ to $\sim6\times10^3\rm \upmu G\,cm^{-3}$, again assuming  \sigmapRM\ of 12 \RMunits. 
A smaller shell radius would likely lead to higher turbulence in the surrounding medium. If the repeating FRB sources are younger compared to (apparent) non-repeaters \citep{Munoz:2020}, a smaller termination shock radius and a higher \sigmapRM\ are expected. 
If non-repeating sources have a larger shell radius, a lower value of $\delta(n_e B_{\rm ||})$ can be expected. Observation of a secular decrease in PRS luminosity for a repeating source would strongly support the age argument of the FRB sources. \citet{Rhodes:2023} report a 30\% decrease in flux for the PRS of FRB20121102A over a 3-year period. However, given the entire variability history of the source's PRS, it remains unclear whether the flux variation is intrinsic and related to a changing termination shock radius or a result of scintillation. Longer-term monitoring is required to elucidate this.

\subsection{Correlation in \sigmaRM and PRS luminosity?}

The observed depolarisation for the repeating FRB sources has been interpreted as being the result of a  complex magneto-ionic local environment. If the large \sigmapRM\ values indeed stem from a complex magnetised and turbulent surrounding medium, one might expect a correlation between \sigmaRM\ and the luminosity of a PRS. We combine our limits on PRS luminosity and measurement of \sigmaRM with published PRS luminosities and \sigmaRM measurements   \citep{Chatterjee:2017, Bhandari:2022, Niu:2022, Feng:2022, Bruni:2024} to explore the correlation between \sigmaRM and PRS luminosity.  
We assume a flat spectrum when comparing PRS luminosities at different frequencies while fitting a simple power law to the \sigmaRM and PRS luminosity of the form 
\begin{equation}
    P_{\rm lum} = A\,\sigma_{\rm RM}^{\alpha},
\end{equation}
where $\alpha$ is the power law index and A is the scaling factor. We use a Gaussian likelihood function of the form 
\begin{equation}
\begin{split}
     \textit{L}(P_{lum, i}| M , \sigma) =  \prod_i^{N_f} \left[\frac{1}{\sqrt{2\pi\sigma^2}}\,\,\exp\left(-\frac{(P_{lum,i}-M)^2}{2\sigma^2}\right)\right],
    \label{eq:posterior_likelihood}
\end{split}
\end{equation}
where $P_{lum, i}$ is the data, M is the model, and $\sigma$ is the standard deviation of the model fit. We only fit three data points for which we have measurements for PRS luminosity, we do not consider upper limits. 

We show the \sigmapRM--PRS luminosity relation in Figure \ref{fig:PRS_sigma_RM} for FRBs given in Table \ref{tab:PRS_sigma_RM}.  We use the upper limits and the measurements of the PRS luminosity to verify any relationship between the FRB \sigmaRM\ and the PRS luminosity. We see an apparent decreasing trend in the \sigmaRM\ and PRS luminosity for the FRBs, indicating a potential correlation between the \sigmaRM\ and L$_{\nu}$. 

There have been a number of models for PRS that predict correlations between \sigmaRM\ and PRS luminosity. 
\cite{Yang:2022} predict the relationship between \sigmaRM\ and the PRS specific luminosity, independent of specific astrophysical model. If we assume the origin of \sigmaRM\ and RM to be in the same medium the luminosity can be estimated to be 
\begin{equation}
L_{\nu} = \frac{64\pi^3 m_e c^2}{27} \left(\frac{\zeta_e \gamma_{\rm th}^2 R^2}{\left(\delta(n_e B_{||})/(n_e B_{||})\right)\,(l_{\rm screen}/R)^{1/2}}\right)\, \sigma_{\rm RM},
\label{eq:luminosity_sigma_RM}
\end{equation}

where $\zeta_{e}$ is the fraction of electrons emitting synchrotron radiation in the GHz band, $\gamma_{\rm th}$ is the Lorentz factor, 
and $R$ is the radius of the depolarising shell \citep{Yang:2022}. 
A higher value of $\delta(n_e B_{\rm ||})$ will lead to a higher \sigmapRM (see Equation 13 in \cite{Yang:2022}) and hence higher specific luminosity. Assuming $R=21$\,pc, \sigmapRM $\sim$12 rad m$^{^2}$, and $(l_{\rm screen}/R)^{1/2}= 0.1$ we find that the fraction of relativistic electrons $\zeta_e \gamma_{\rm th}^2$ to be on the order of $<$10$^6$ to match the 3-$\sigma$ upper limit of PRS luminosity of FRB~20230526A. This suggests a lower fraction of relativistic electrons ($\zeta_e \gamma_{\rm th}^2$) in the immediate surrounding local environment for this (apparent) non-repeater, compared to active repeating FRB sources. However, we caution the reader that a larger sample of FRB-PRS associations is still needed to conclusively derive any relationship between \sigmaRM\ and L$_{\nu}$. Hence, sensitive searches for PRS and \sigmaRM associated with FRB sources will be crucial in our understanding of the evolution and progenitor environment of FRBs.


We note that the lack of detection of PRS in some of the repeating FRB sources which have shown measurable depolarisation \cite[e.g., FRBs~20180301A, 20180916B, 20190303A;][]{Feng:2022} indicates that detectable PRS emission association is not a necessary condition for depolarisation. These FRBs have lower values of \sigmapRM\ and RM than those that show PRS. However, observational biases in the search for PRS cannot yet be ruled out for FRBs showing depolarisation. For example, FRBs~20121102A and 20190520B had relatively flatter spectral index for their PRSs \citep{Chatterjee:2017, Niu:2022}, but FRB~20201124A showed an inverted spectral index \citep[brighter at higher frequencies;][]{Bruni:2024}.
This motivates further sensitive high frequency $>20$\,GHz observations to search for the presence of PRS.



\section{Conclusions}
\label{sec:conslusion}
Our search for and study of spectral depolarisation  in an extended  ASKAP ICS FRB sample shows a diverse and intriguing polarisation behaviour of FRB sources. From this analysis we observe the following:

\begin{itemize}
    \item[1.] We report a detection of depolarisation for FRB~20230526A, the first for any apparently non-repeating FRB source.  We also find modest evidence for depolarisation in FRB~20210320C.  For the remaining $10$ FRBs in our sample we find no evidence for depolarisation.  
    
    \item[2.] The 10 out of 12 sources in this work that do not have detectable \sigmapRM\ are inconsistent with the \sigmapRM, RM, and \taus\ relationships established for repeaters by \citet{Feng:2022}. Our finding is consistent with that reported for a smaller sample by \cite{Uttarkar:2023}.
    The measured RM are consistent with  RM-\sigmapRM relationship only if  \sigmaRM\ is $\sim$2 orders of magnitude lower than our upper limits. In any case our \sigmaRM\ upper limits are inconsistent with the \taus\ and \sigmapRM\ trend reported by \cite{Feng:2022}.

    \item[3.] FRBs 20210320C and 20230526A fall within 1-$\sigma$ of the relationships between \sigmapRM-RM and \sigmapRM-\taus\ found by \citet{Feng:2022}, indicating that depolarisation, scatter broadening, and RM are consistent with originating in the source environment.

    \item[4.] The location of FRB 20230526A was observed with ATCA to search for the presence of a persistent radio source. However, we do not detect any putative continuum source co-located with the FRB localisation position. We put an upper limit of $\sim$20 $\upmu$Jy or $<0.11\times10^{29}$ erg s$^{-1}$ Hz$^{-1}$ for any PRS associated with the source.

    \item[5.] We explore the relationship between PRS and \sigmaRM based on the relationship given by \cite{Yang:2022}. We see an apparent decrease in \sigmaRM with a decrease in the luminosity of the PRS associated with the FRBs. However, a larger sample of FRBs associated with PRS is crucial to conclude on the correlation between \sigmaRM and the spectral luminosity of PRS.

\end{itemize}

The investigation of depolarisation in this extended dataset from the ASKAP ICS survey suggests that the repeating and non-repeating FRB sources form a continuum from a single population. Some non-repeating FRB sources have environments more similar to, but more moderate than, repeating sources, indicating an evolving surrounding environment of the FRB sources, where the complex circumburst environment evolves into a less dense environment and PRS emission fades away.
Eventually the effects of the environment cannot be seen, and other media (such as the bulk ISM from the host galaxy) contribute to measured FRB RM and scatter broadening. 
In this scenario the FRBs sources remain active occasionally producing FRBs within a placid environment. 
These old FRBs are some of the most luminous and distant, thus they can be used as clean probes of the intergalactic medium of the high-redshift Universe.   


\section*{Acknowledgements}

PAU, RMS, and YW acknowledge support through Australia Research Council Future Fellowship FT190100155. RMS, ATD, and YW  acknowledge support through Australian Research Council Discovery Project DP220102305.
MG is supported by the Australian Government through the Australian Research Council’s Discovery Projects funding scheme (DP210102103), and through UK STFC Grant ST/Y001117/1. MG acknowledges support from the Inter-University Institute for Data Intensive Astronomy (IDIA). IDIA is a partnership of the University of Cape Town, the University of Pretoria and the University of the Western Cape. 

This scientific work uses data obtained from Inyarrimanha Ilgari Bundara / the Murchison Radio-astronomy Observatory. We acknowledge the Wajarri Yamaji People as the Traditional Owners and native title holders of the Observatory site. CSIRO’s ASKAP radio telescope is part of the Australia Telescope National Facility (https://ror.org/05qajvd42). Operation of ASKAP is funded by the Australian Government with support from the National Collaborative Research Infrastructure Strategy. ASKAP uses the resources of the Pawsey Supercomputing Research Centre. Establishment of ASKAP, Inyarrimanha Ilgari Bundara, the CSIRO Murchison Radio-astronomy Observatory and the Pawsey Supercomputing Research Centre are initiatives of the Australian Government, with support from the Government of Western Australia and the Science and Industry Endowment Fund.

The Australia Telescope Compact Array is part of the Australia Telescope National Facility (https://ror.org/05qajvd42) which is funded by the Australian Government for operation as a National Facility managed by CSIRO. We acknowledge the Gomeroi people as the Traditional Owners of the Observatory site. 

Murriyang, the Parkes radio telescope, is part of the Australia Telescope National Facility (https://ror.org/05qajvd42) which is funded by the Australian Government for operation as a National Facility managed by CSIRO. We acknowledge the Wiradjuri people as the Traditional Owners of the Observatory site

Based on observations made with ESO Telescopes at the La Silla Paranal Observatory under programme ID 108.21ZF.
A.C.G. and the Fong Group at Northwestern acknowledges support by the National Science Foundation under grant Nos. AST-1909358, AST-2308182 and CAREER grant No. AST-2047919.
A.C.G. acknowledges support from NSF grants AST-1911140, AST-1910471 and AST-2206490 as a member of the Fast and Fortunate for FRB Follow-up team.

This work was performed on the OzSTAR national facility at Swinburne University of Technology. This work makes use of OzSTAR supercomputing facility. The OzSTAR program receives funding in part from the Astronomy National Collaborative Research Infrastructure Strategy (NCRIS) allocation provided by the Australian Government, and from the Victorian Higher Education State Investment Fund (VHESIF) provided by the Victorian Government.

\section*{Data Availability}

The code base and processed data used in this work is available at \url{github.com/pavanuttarkar/depol_extended}.



\bibliographystyle{mnras}
\bibliography{example} 




\appendix

\section{FRB depolarisation measurements}
In Figure  \ref{fig:depolarisation_model_apdx}, we show the depolarisation spectra of FRBs for which we did not find any evidence of depolarisation, i.e., the FRBs for which the Bayes factor (BF) is $\leq$1 The linear polarisation fraction for all the FRBs were calculated with a spectral resolution of 20.978 MHz. We use simple boxcar for temporal averaging of the pulse for FRBs 20220501C, 20221106A, and 20230718A. A match filter was used for all of the remaining FRBs.   

\begin{figure*}
    \begin{subfigure}[t]{.4\linewidth}
        \centering\includegraphics[width=1\linewidth]{DEPOLARISATION_PROFILES_NEWDATA/20191228A.pdf}
        \caption{}
    \end{subfigure}
    \begin{subfigure}[t]{.4\linewidth}
        \centering\includegraphics[width=1\linewidth]{DEPOLARISATION_PROFILES_NEWDATA/20200430A.pdf}
        \caption{}
    \end{subfigure}\\
    \begin{subfigure}[t]{.4\linewidth}
        \centering\includegraphics[width=1\linewidth]{DEPOLARISATION_PROFILES_NEWDATA/20210117A.pdf}
        \caption{}
    \end{subfigure}  
    \begin{subfigure}[t]{.4\linewidth}
        \centering\includegraphics[width=1\linewidth]{DEPOLARISATION_PROFILES_NEWDATA/20210407E.pdf}
        \caption{}
    \end{subfigure}  
    \begin{subfigure}[t]{.4\linewidth}
        \centering\includegraphics[width=1\linewidth]{DEPOLARISATION_PROFILES_NEWDATA/20210912A.pdf}
        \caption{}
    \end{subfigure}
    \begin{subfigure}[t]{.4\linewidth}
        \centering\includegraphics[width=1\linewidth]{DEPOLARISATION_PROFILES_NEWDATA/20220501C.pdf}
        \caption{}
    \end{subfigure}  
    
\end{figure*}

\begin{figure*}
 \ContinuedFloat
    \begin{subfigure}[t]{.4\linewidth}
        \centering\includegraphics[width=1\linewidth]{DEPOLARISATION_PROFILES_NEWDATA/20220610A.pdf}
        \caption{}
    \end{subfigure}  
    \begin{subfigure}[t]{.4\linewidth}
        \centering\includegraphics[width=1\linewidth]{DEPOLARISATION_PROFILES_NEWDATA/20220725A.pdf}
        \caption{}
    \end{subfigure} 
        \begin{subfigure}[t]{.4\linewidth}
        \centering\includegraphics[width=1\linewidth]{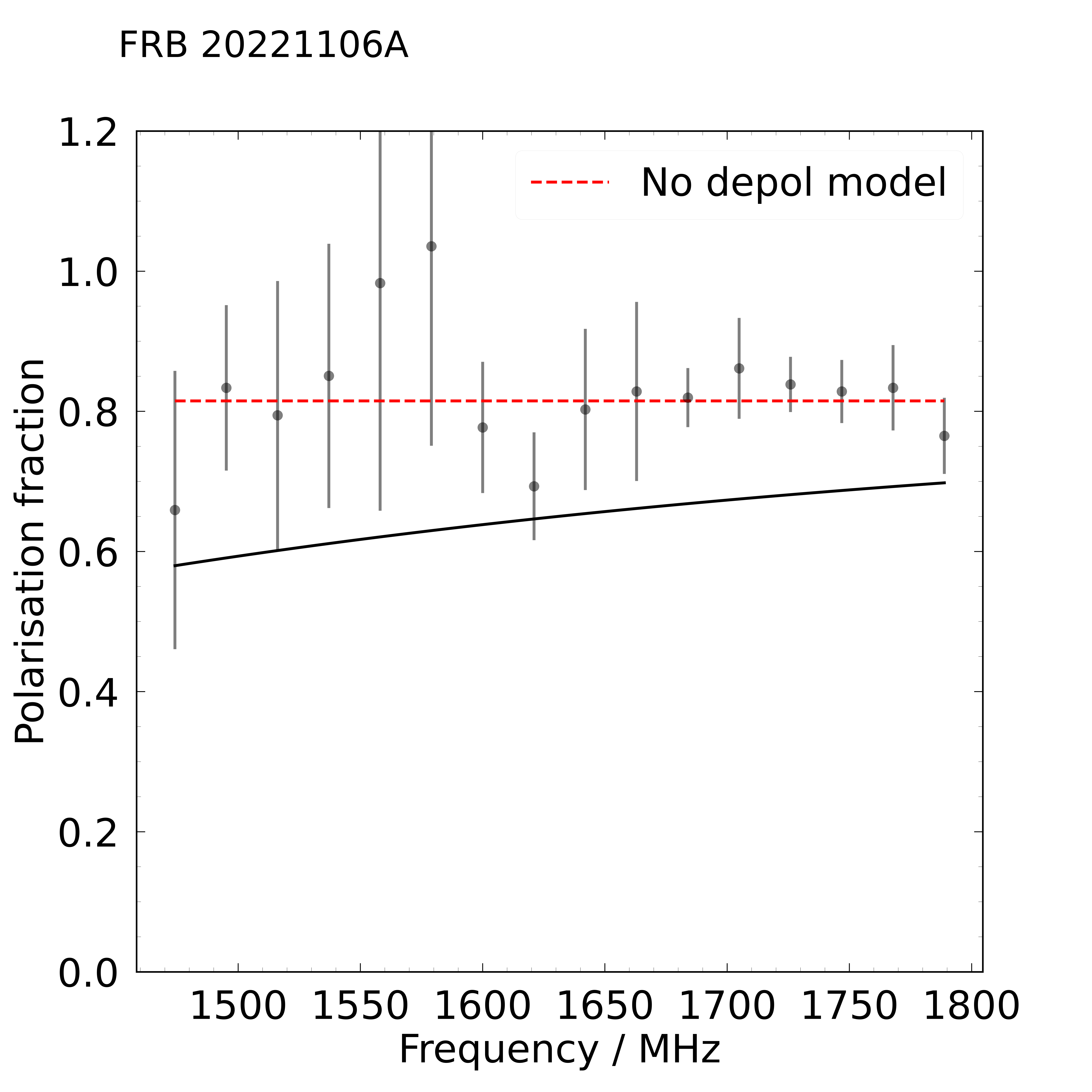}
        \caption{}
    \end{subfigure}
    \begin{subfigure}[t]{.4\linewidth}
        \centering\includegraphics[width=1\linewidth]{DEPOLARISATION_PROFILES_NEWDATA/20230718A.pdf}
        \caption{}
    \end{subfigure}

    \caption{Linear polarisation fraction spectra. We show the FRBs that do not show evidence for depolarisation. The figures show the spectral depolarisation profiles for FRBs showing weak Bayes evidence for modified Burn's law. We show the favoured model for each burst in dashed line. The 95\% upper limit from the modified Burns law is shown for all the bursts in solid black line.}
    \label{fig:depolarisation_model_apdx}
\end{figure*}


\bsp	
\label{lastpage}
\end{document}